\documentclass[aps,prc,twocolumn,floatfix,superscriptaddress,amsmath,amssymb,nofootinbib]{revtex4-2}

\usepackage[utf8]{inputenc} 
\usepackage[T1]{fontenc}
\usepackage{dcolumn}
\usepackage{braket}
\usepackage{float}
\usepackage{graphicx}
\usepackage[colorlinks=true,linkcolor=blue,citecolor=blue,urlcolor=blue]{hyperref}
\usepackage{orcidlink}

\graphicspath{{.}{./figures/}}

\begin{document}

\title{Structure of the doubly magic nuclei $^{208}$Pb and $^{266}$Pb from \textit{ab initio} computations}


\author{Francesca~Bonaiti\orcidlink{0000-0002-3926-1609}}
\email{bonaiti@frib.msu.edu}
\affiliation{Facility for Rare Isotope Beams, Michigan State University, East Lansing, Michigan 48824, USA}
\affiliation{Physics Division, Oak Ridge National Laboratory, Oak Ridge, Tennessee 37831, USA}

\author{Gaute~Hagen\orcidlink{0000-0001-6019-1687}}
\affiliation{Physics Division, Oak Ridge National Laboratory, Oak Ridge, Tennessee 37831, USA}
\affiliation{Department of Physics and Astronomy, University of Tennessee, Knoxville, Tennessee 37996, USA}

\author{Thomas~Papenbrock\orcidlink{0000-0001-8733-2849}}
\affiliation{Department of Physics and Astronomy, University of Tennessee, Knoxville, Tennessee 37996, USA}
\affiliation{Physics Division, Oak Ridge National Laboratory, Oak Ridge, Tennessee 37831, USA}

\begin{abstract}
Theoretical studies indicate that the superheavy neutron-rich nucleus  $^{266}_{\ 82}$Pb$_{184}$ 
is doubly magic and at the neutron drip line. While its density distributions and single-particle energies have been computed, the structure of this nucleus is yet unknown. We perform {\it ab initio} computations of $^{266}$Pb using an interaction from an effective field theory of quantum chromodynamics tuned only on properties of nuclei with $A\leq 4$. We validate our theoretical framework by computing the first $2^+$ and $3^-$ excited states of $^{208}$Pb, finding agreement with experimental data. We confirm that $^{266}$Pb is doubly magic and show that its $3^-$ state, located below the $2^+$ state, exhibits an excitation gap of 2.6 MeV with respect to the ground state. Our calculations also suggest that this nucleus is at the neutron drip line.    
\end{abstract}
\maketitle

\emph{Introduction.---} 
Doubly magic nuclei exhibit filled spherical shells for protons and neutrons~\cite{mayer1955}. They are more strongly bound, more compact, and more difficult to excite than their neighbors and serve as cornerstones for our understanding of nuclei in entire regions of the Segr{\`e} chart. Established doubly magic nuclei are  $^4$He, $^{16}$O, $^{40,48}$Ca, and the rare isotopes $^{78}$Ni~\cite{taniuchi2019} and $^{132}$Sn~\cite{jones2010}. At present, this list terminates with the heaviest known $\beta$-stable nucleus $^{208}$Pb. 

Since the 1960s, extrapolations to the next shell closures beyond $^{208}$Pb fueled the search for an ``island of stability" of long-lived superheavy elements and its location in the nuclear chart. In particular, theorists identified neutron number $N=184$ as another traditional shell closure~\cite{myers1966,sobieczewski1966,meldner1967,strutinsky1967,nilsson1969}; for more recent studies see, e.g., Refs.~\cite{cwiok1996,bender1999,bender2003,giuliani2019}. Computations of separation energies and effective single-particle energies also identified $^{266}$Pb (with $Z = 82$ and $N = 184$) as a doubly magic nucleus~\cite{nakada2014,li2014,afanasjev2015,agbemava2015}. Bulk properties of this nucleus (such as density distributions and strengths functions)  were computed in Refs.~\cite{Li1998,centelles2010,kim2022}. Surveys of the nuclear landscape, based non-relativistic and covariant density functionals, leave little doubt that $^{266}$Pb is at the neutron drip line~\cite{meng2002,stoitsov2003,erler2012,afanasjev2013}. Besides these few results, little else is known about $^{266}$Pb and its structure is anyone's guess.

While superheavy nuclei have thus far been studied in mean-field-based approaches, recent years have seen a remarkable progress in \textit{ab initio} nuclear theory~\cite{hergert2020,Ekstrom:2022yea}. \textit{Ab initio} computations aim to solve the nuclear quantum many-body problem at the highest practical resolution scale. They use protons and neutrons as degrees of freedom, employ nuclear interactions rooted in the fundamental theory of quantum chromodynamics  via chiral effective field theory ~\cite{weinberg1990,epelbaum2009,machleidt2011}, and use systematically improvable many-body methods~\cite{hagen2014,dickhoff2004,lee2009,hergert2016,stroberg2019}.
\textit{Ab initio} calculations recently reached the heavy nucleus $^{208}$Pb~\cite{hu2022,hebeler2023,arthuis2024}, opening the door to first-principles studies of superheavy isotopes.

In this Letter we reveal the structure of $^{266}$Pb via \textit{ab initio} computations of its low-lying excited states. 
This is not only of academic interest. Experimenters and theorists have pushed their studies beyond the $N=126$ shell closure
~\cite{caballero2016,houda2021,yuan2022,yeung2024,lica2025}, see Ref.~\cite{kiss2024} for a recent review. In lead nuclei $N=134$ has been reached~\cite{pfutzner1998,valientedobon2021,gottardo2012}. While this is still 50 neutrons away from $^{266}$Pb, \textit{ab initio} theory can now compute the structure of this nucleus. 

\begin{figure}[hbt]
    \centering
    \includegraphics[width=0.49\textwidth]{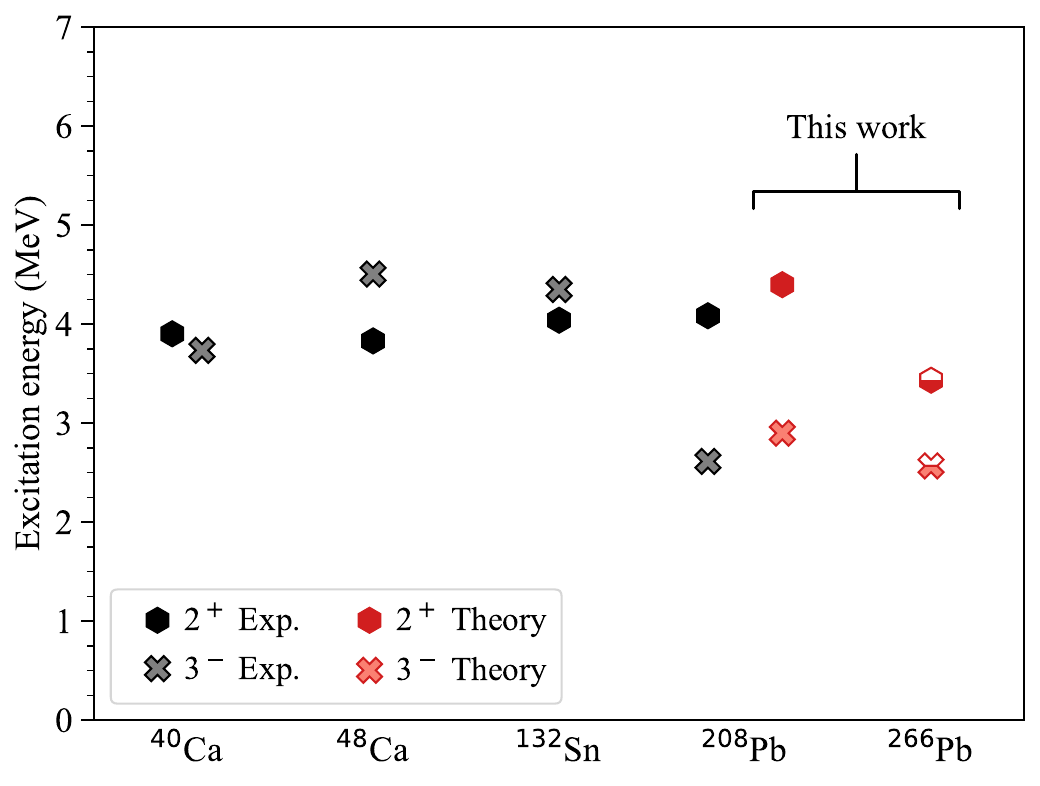}
    \caption{Excitation energy of the first $2^{+}$ and $3^-$ excited states of $^{208}$Pb and $^{266}$Pb from the \textit{ab initio} computations of this work, based on the chiral interaction 1.8/2.0 (EM)~\cite{hebeler2011}, in comparison to experimental data for $^{40,48}$Ca, $^{132}$Sn, $^{208}$Pb from Ref.~\cite{nudat}. Predictions for $^{266}$Pb, shown with half-filled markers, are derived from calculations performed in a reduced model space, adjusted by an estimate of the contribution expected from larger basis sizes. See text for details.} 
    \label{fig:summary}
\end{figure}

Our main results are presented in Fig.~\ref{fig:summary}, where we show the energies of the first excited $J^\pi=2^+$ and $3^-$ states in $^{208}$Pb and $^{266}$Pb and compare them to other doubly magic nuclei.  
Our theoretical results are in good agreement with experiment for $^{208}$Pb. For $^{266}$Pb we find a similar energy gap between the excited $3^-$ state and the ground state; this identifies $^{266}$Pb as a doubly magic nucleus. However, the $2^+$ state in $^{266}$Pb is about 1~MeV lower in energy than in $^{208}$Pb. 
In what follows we describe how we computed these results and also present results for neighboring nuclei. We also assess the uncertainties associated with the results shown in Fig.~\ref{fig:summary} by studying the sensitivity of the energies with respect to model-space parameters.

\emph{Hamiltonian and model space.---} The starting point of our coupled-cluster computations is the intrinsic nuclear Hamiltonian
\begin{equation}
    \hat{H} = \sum_{i<j} \left(\frac{(\mathbf{p}_i -\mathbf{p}_j)^2}{2mA} + \hat{V}^{NN}_{ij}\right) + \sum_{i<j<k} \hat{V}^{3N}_{ijk}. 
    \label{hamiltonian}
\end{equation}
It consists of the intrinsic kinetic energy (i.e. the kinetic energy of the center of mass is removed), the nucleon-nucleon interaction, and three-nucleon potential. 
In this work we employ the chiral interaction 1.8/2.0 (EM) from Ref.~\cite{hebeler2011}. It is derived by applying similarity-renormalization-group~\cite{bogner2007} transformations to the high-precision nucleon-nucleon potential by \textcite{entem2003} and also includes the leading three-nucleon forces from chiral effective field theory~\cite{epelbaum2006}. The low-energy constants of the latter are optimized on ground-state properties of nuclei with mass numbers $A = 3, 4$. 
This interaction yields accurate ground-state energies and excitation spectra of the closed-shell nuclei $^4$He, $^{16}$O, $^{40,48}$Ca, and $^{78}$Ni~\cite{hagen2016b,simonis2016,simonis2017}, and it matches large-scale shell model results for $^{100}$Sn~\cite{morris2018}. These benchmarks on binding energies and spectra in different regions of the nuclear chart make it a suitable choice for the goals of this paper.

We employ a Hartree-Fock basis computed in a spherical harmonic oscillator basis with single-particle excitations up to $N_{\rm max}\hbar\Omega$. Our largest model spaces use up to 15 major harmonic oscillator shells, i.e. $N_{\rm max}=14$. We compute $^{208, 209}$Pb and $^{209}$Bi at a harmonic oscillator frequency of $\hbar\Omega = 12$~MeV and use $\hbar\Omega = 10$~MeV for $^{266, 267}$Pb and $^{267}$Bi. These values roughly yield energy minima in the largest model-space size under consideration. We employ three-nucleon potentials in the normal-ordered two-body approximation~\cite{hagen2007a,roth2012,takayuki2022,rothman2025} and compute matrix elements with the codes by \textcite{miyagi2023,stroberg_code}. 
In these computations we adopt an energy cut $E_{3,{\rm max}} = 28 \hbar\Omega$ on three-body matrix elements~\cite{takayuki2022}, which ensures sufficiently well converged ground-state energies and excited spectra for $^{208}$Pb and $^{266}$Pb.

\emph{Method.---} We solve the quantum many-body problem with the coupled-cluster method~\cite{coester1960,kuemmel1978,bartlett2007} in an angular-momentum-coupled scheme~\cite{hagen2008,hagen2014}. Coupled-cluster theory is based on an exponential ansatz for the nuclear ground-state wave function $\ket{\Psi_0} = e^T \ket{\Phi_0}$, where the reference state $\ket{\Phi_0}$ is typically obtained from the Hartree-Fock solution, and the cluster operator $T = T_1 + T_2 + \dots T_n+ \ldots + T_A$ is a sum of $n$-particle--$n$-hole excitation operators. In the coupled-cluster singles and doubles (CCSD) approximation one truncates $T=T_1+T_2$~\cite{bartlett2007}. This approximation  captures about 90\% of the ground-state correlation energy in closed shell systems~\cite{bartlett2007,hagen2009b,sun2022}. For $^{208,266}$Pb, we compute the similarity-transformed Hamiltonian 
\begin{equation}
    \overline{H}\equiv \exp{(-T_1-T_2)}H\exp{(T_1+T_2)}
\end{equation}
 in the CCSD approximation. Excited states in $^{208,266}$Pb and in the neighboring (particle attached) nuclei $^{209,267}$Pb are computed as generalized exitations of $\overline{H}$. These are accurate for excited states that are dominated by 1-particle--1-hole and 1-particle excitations in the closed-shell and the particle-attached nuclei, respectively.
To compute low-lying excited states of $^{208,266}$Pb, we use the equation-of-motion (EOM) coupled-cluster method in the EOM-CCSD approximation~\cite{stanton1993}.
To estimate the impact of higher-order many-body correlations, we also use perturbative 3-particle--3-hole excitations in the EOM-CCSD(T) approach~\cite{watts1995}. This method offers a cost-effective and accurate improvement over EOM-CCSD~\cite{watson2013,hagen2016b,morris2018}. 

To investigate shell structure, we also study the nuclei $^{209,267}$Pb and $^{209,267}$Bi that differ from $^{208,266}$Pb by attaching a single neutron or proton, respectively. We compute the spectra of these nuclei with the particle-attached EOM-CCSD (PA-EOM-CCSD) technique~\cite{gour2005,hagen2009b,hagen2010a}. This approach involves acting with an effective one-particle creation operator onto the ground-state of the neighboring closed-shell system. We truncate the creation operator expansion at 2-particle--1-hole (2p-1h) level, which is accurate for states with a dominant single-particle character. To assess how controlled this approximation is, we compute perturbative 3-particle--2-hole (3p-2h) corrections as in Ref.~\cite{morris2018} for $^{209}$Pb (which we denote as PA-EOM-CCSD(3p-2h)$_{\rm pert}$).

\emph{Validation in $^{208}$Pb.---} The nucleus $^{208}$Pb and its low-energy spectrum are very well known~\cite{heusler2016}. To validate our approach we compute the ground-state energy and the low-lying $3^-$ and $2^+$ excited states of $^{208}$Pb. The CCSD ground-state energy is $-1630$~MeV. The energy difference between the two largest model spaces  ($N_{\rm max} = 12$ and $N_{\rm max} = 14$) is about  20~MeV, corresponding to about $1\%$ of the total binding energy. While this level of precision is typical for \textit{ab initio} computations, it is much lower than nuclear mass models which have a root-mean-square deviation of the order of 1~MeV~\cite{goriely2009,moller2006,kortelainen2010,zurek2024}. The 1.8/2.0 (EM) result for the CCSD ground-state energy is in agreement with the results from a set of non-implausible interactions~\cite{hu2022}, while it lies $2\%$ higher than the in-medium similarity renormalization group calculations in the IMSRG(2) truncation~\cite{hebeler2023}. This is due to a different counting of many-body contributions in CCSD and IMSRG(2)~\cite{hergert2016}. Considering that triples corrections typically amount to $10\%$ of the CCSD correlation energy, we estimate that the 1.8/2.0 (EM) interaction would lead to a $2\%$ overbinding in $^{208}$Pb with respect to the experimental value of $-1636.336$ MeV~\cite{wang2021}.

\begin{figure}[t!]
    \centering
    \includegraphics[width=0.49\textwidth]{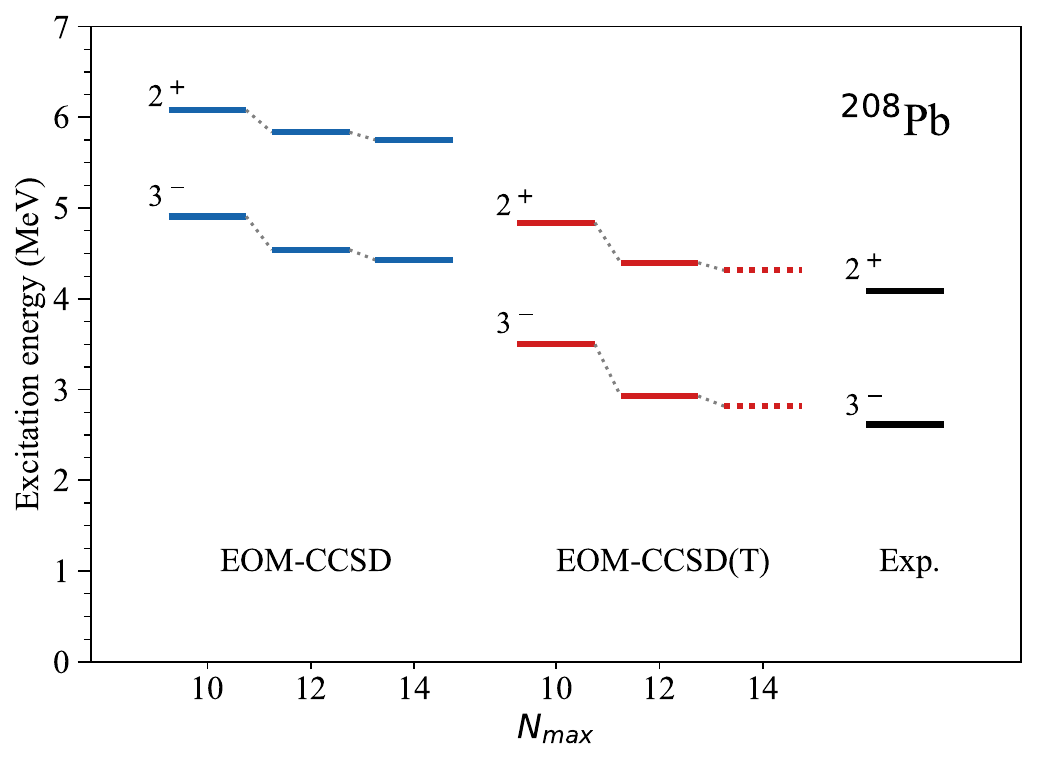}
    \caption{Model space convergence of the first $2^+$ and $3^-$ excited states of $^{208}$Pb calculated with the EOM-CCSD and EOM-CCSD(T) approaches employing the 1.8/2.0 (EM) interaction. The triples result for $N_{\rm max} = 14$ is shown as a dotted line because it is based on adding to the $N_{\rm max} = 12$ result the additional $N_{\rm max} = 14$ energy contribution obtained at the EOM-CCSD level. A comparison with experimental data from Ref.~\cite{nudat} is also presented.}
    \label{fig:spectrum208}
\end{figure}

Let us consider the low-lying spectrum of $^{208}$Pb. The first excited state in $^{208}$Pb has spin/parity $J^\pi=3^-$. It is interpreted as a collective octupole vibration~\cite{goutte1980,spear1983}. In a shell model picture the low-lying $2^+$ state is associated with single-neutron excitations, involving a $1i_{13/2} \rightarrow 2g_{9/2}$ transition~\cite{schramm1997,heusler2013}. We also carried out calculations for the first $4^-$ state. However, the results showed poor model space convergence, and we therefore omit them from the present analysis. We compute the excitation energy of these states in an \textit{ab initio} framework. The results of our EOM-CCSD and EOM-CCSD(T) computations for the $3^-$ and $2^+$ excited states of $^{208}$Pb are shown in Fig.~\ref{fig:spectrum208} and compared to data~\cite{nudat}. 

At $N_{\rm max} = 14$, excitation energies at the EOM-CCSD level are practically converged with respect to the model space size. The difference between $N_{\rm max} = 12$ and $N_{\rm max} = 14$ amounts to only 80 and 110~keV for the $2^+$ and $3^-$ states, respectively. Due to the computational cost, we perform EOM-CCSD(T) calculations only for $N_{\rm max} = 10, 12$. In Fig.~\ref{fig:summary}, we report the result of our EOM-CCSD(T) computation at the largest available model space size of $N_{\rm max} =12$. We can also estimate the $N_{\rm max} = 14$ value, shown by with a dotted line in Fig.~\ref{fig:spectrum208}, by adding to the $N_{\rm max} = 12$ result the variation in excitation energy between the $N_{\rm max} = 12$ and $N_{\rm max} = 14$ energies observed at the EOM-CCSD level. Triples contributions lower excitation energies of around $1.5$~MeV, moving our results closer to experiment. This is consistent with previous analyses of the effect of 3-particle--3-hole excitations on the $2^+$ state in $^{48}$Ca and $^{78}$Ni~\cite{hagen2016b}, and on the low-lying spectrum of $^{100}$Sn~\cite{morris2018}. Our EOM-CCSD(T) results are in good agreement with experiment, and this gives us confidence in extending our coupled-cluster calculations based the 1.8/2.0 (EM) interaction to $A>208$.

\emph{Excited states in $^{209}$Bi and $^{209}$Pb.---} To get a picture about the evolution of  shell structure beyond $^{208}$Pb, we also  computed the spectra of the neighboring nuclei $^{209}$Bi and $^{209}$Pb. 
(Results for the excited states of $^{209}$Bi are shown in the Supplemental Material.)

\begin{figure}[htb]
    \centering
    \includegraphics[width=0.49\textwidth]{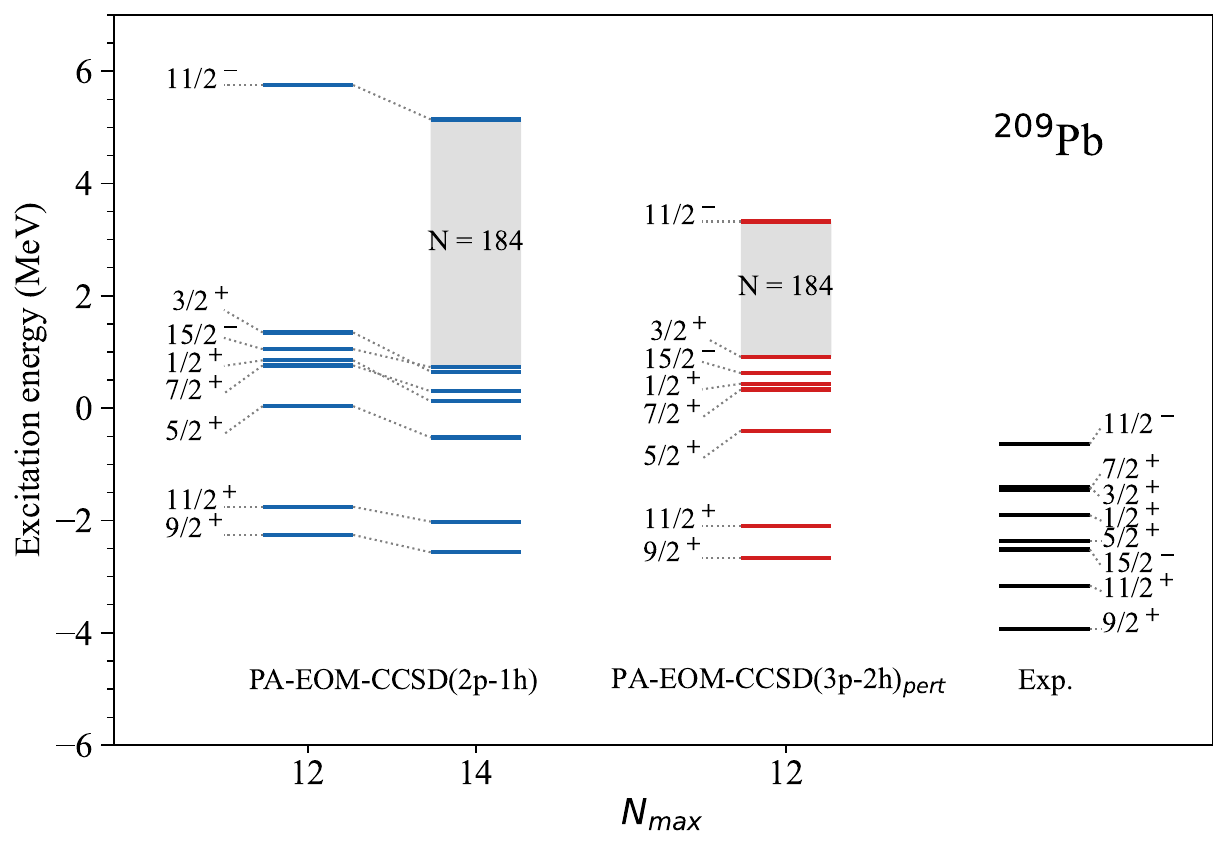}
    \caption{Excited states of $^{209}$Pb computed with the PA-EOM-CCSD(2p-1h) and PA-EOM-CCSD(3p-2h)$_{\rm pert}$ approaches employing the 1.8/2.0 (EM) chiral interaction. The shell gap corresponding to $N=184$ is highlighted in gray. Comparison with the experimental data of Ref.~\cite{nudat} is also shown.}
    \label{fig:spectrum209}
\end{figure}

Let us focus  on  $^{209}$Pb. Our PA-EOM-CCSD results at the 2p-1h and 3p-2h truncation levels are shown in Fig.~\ref{fig:spectrum209} and compared to data~\cite{nudat}. All energies are with respect to the ground state of $^{208}$Pb. We employed $N_{\rm max} = 12, 14$ for the PA-EOM-CCSD results at the 2p-1h level, while due to computational cost we are limited to $N_{\rm max} = 12$ in calculating 3p-2h corrections.
The computed spectra are inflated with respect to data, although perturbative 3p-2h corrections somewhat improve the picture. 
At the 2p-1h level, increasing the model space size decreases the excitation energy of an amount ranging between 0.3 and 0.7~MeV and the ordering of the states also changes. The inclusion of 3p-2h excitations leads to a reduction in excitation energy of around 0.4~MeV, except for the case of the $11/2^-$ state, where 3p-2h contributions amount to 2.4~MeV. This can be understood by looking at the single-particle character of these states. The one-particle--zero-hole (1p-0h) contribution to the norm of the wavefunction amounts to $90\%$ for all the computed states of $^{209}$Pb  except for the $11/2^-$ state, where it is only $70\%$. This suggests that higher-order correlations are needed to provide an accurate description of this state. 

Despite the large effect of 3p-2h excitations on the $11/2^-$ state, we observe a pronounced gap between the $3/2^{+}$ and the $11/2^{-}$ states, suggesting that $N = 184$ emerges as a magic neutron number. This \textit{ab initio}  result agrees with results from energy density functionals~\cite{cwiok1996,bender2001,agbemava2015,giuliani2019} and supports that $^{266}$Pb is a doubly magic nucleus. We note that in order to rigorously quantify a shell gap in the effective single particle spectrum we would need to diagonalize the centroid matrix as discussed in Ref.~\cite{duguet2012}. In contrast to our computed energy spectra these effective single particle energies are model-dependent and non-observable~\cite{duguet2012,duguet2015b}.

\emph{Structure of $^{266}$Pb.---} The CCSD approximation yields a ground-state energy of $-1730$~MeV, with a $3\%$ uncertainty due to the dependence on the model-space size. This corresponds to a binding energy per nucleon of around $6.5$~MeV, notably smaller than the $7.867$~MeV for $^{208}$Pb~\cite{wang2021}. It is also lower than the relativistic mean-field result of $6.85$ MeV from Ref.~\cite{centelles2010}. However, in that work the same functionals lead to overbinding in $^{208}$Pb~\cite{centelles2010}.

\begin{figure}[htb]
    \centering
    \includegraphics[width=0.49\textwidth]{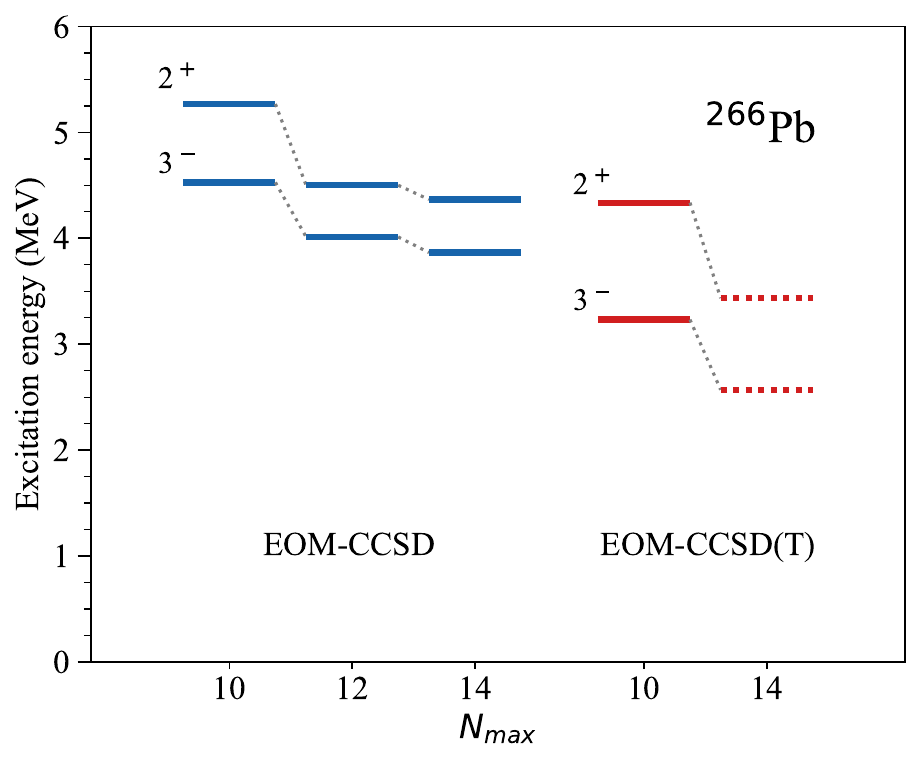}
    \caption{Model space convergence of the first $2^+$ and $3^-$ excited states of $^{266}$Pb computed with the EOM-CCSD and EOM-CCSD(T) approaches employing the 1.8/2.0 (EM) interaction. The triples result for $N_{\rm max} = 14$ is shown with a dotted line as it is based on adding to the $N_{\rm max} = 10$ result the difference between the $N_{\rm max} = 10$ and $N_{\rm max} = 14$ results obtained at the EOM-CCSD level.}
    \label{fig:spectrum266}
\end{figure}

Figure~\ref{fig:spectrum266} shows our prediction for the excitation energies of the first $3^{-}$ and $2^{+}$ states.
In a model space with $N_{\rm max} = 14$, EOM-CCSD results are converged within $140$~keV. For $N_{\rm max} = 10$, we are able to compute also 3p-3h contributions. In this case, triples lead to a reduction in excitation energy of 0.9 and 1.3~MeV for the $2^+$ and $3^-$ states, respectively, in line with the systematics observed in $^{208}$Pb (see Fig.~\ref{fig:spectrum208}), in $^{48}$Ca and $^{78}$Ni (see Ref.~\cite{hagen2016b}), and in $^{100}$Sn (see Ref.~\cite{morris2018}). We can estimate the converged result by adding to the $N_{\rm max} = 10$ result the difference between the $N_{\rm max} = 10$ and $N_{\rm max} = 14$ EOM-CCSD values. This prediction is shown with a half-filled marker in Fig.~\ref{fig:summary} and with a dotted line in Fig.~\ref{fig:spectrum266}. According to this assessment, the $2^+$ state is at an excitation energy of $3.4$~MeV, higher than that of the neutron-rich doubly-magic nucleus $^{78}$Ni~\cite{taniuchi2019}. Our predictions thus identify $^{266}$Pb as a doubly-magic nucleus. This result is consistent with the shell gaps from mean-field calculations~\cite{nakada2014,li2014,afanasjev2015,kim2022,agbemava2015,giuliani2019} but obtained here directly from observables. 

\emph{Excited spectra of $^{267}$Bi and $^{267}$Pb.---} Starting from $^{266}$Pb, we compute low-lying states in the neighboring nuclei $^{267}$Bi and $^{267}$Pb in the PA-EOM-CCSD approach. We present the low-lying spectrum of $^{267}$Bi in the Supplemental Material.  

Figure~\ref{fig:spectrum267} shows our predictions for $^{267}$Pb using the PA-EOM-CCSD(2p-1h) approach. Energies are shown with respect to the $^{266}$Pb ground state.
\begin{figure}[htb]
    \centering
    \includegraphics[width=0.49\textwidth]{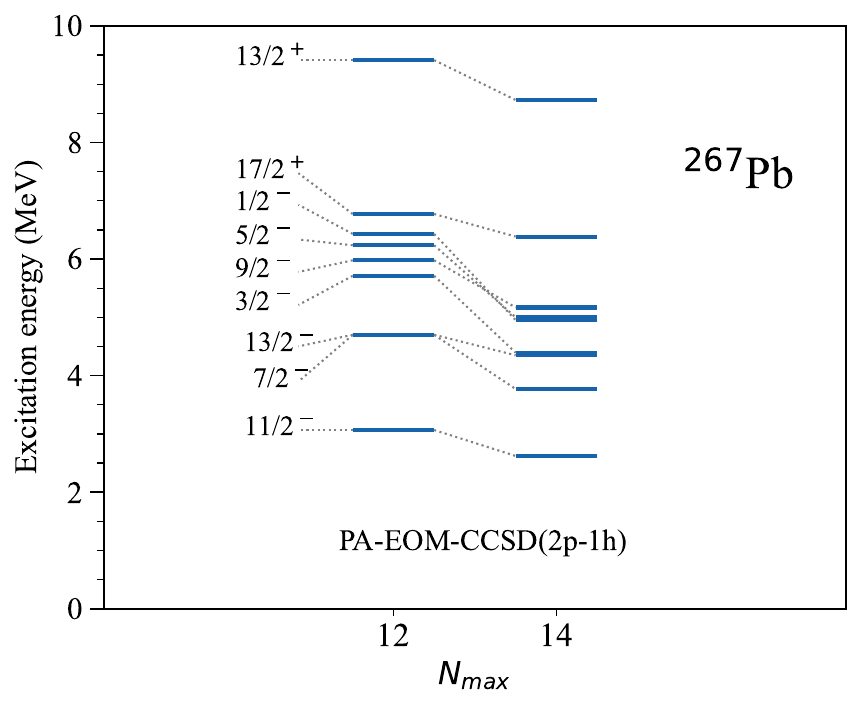}
    \caption{Excited states of $^{267}$Pb computed with the PA-EOM-CCSD(2p-1h) approach employing the 1.8/2.0 (EM) chiral interaction.}
    \label{fig:spectrum267}
\end{figure}
We find that all states calculated with the 2-particle--1-hole approximation are unbound with respect to the ground state of $^{266}$Pb. We note that orbital angular momenta of the lowest lying states in $^{267}$Pb are finite and that the angular momentum barrier prevents the formation of halo states. Thus, the impact of the continuum -- which we neglected -- is expected to be small. As for $^{209}$Pb, we observe a slow convergence of energies with respect to increasing model space size. The neutron separation energy $S_n\approx -3$~MeV of the $11/2^-$ state in $^{267}$Pb suggests that $^{266}$Pb is located at the neutron dripline, assuming that all lighter isotopes of lead are bound. 
Our \textit{ab initio} identification of $^{266}$Pb as a dripline nucleus agrees with the results from non-relativistic~\cite{erler2012,kim2022} and covariant~\cite{meng2002,zhang2003,afanasjev2005} energy density functionals.

We finally note that there is no energy gap clearly identifiable in the spectrum of $^{267}$Pb, except for a separation of about $2.3$~MeV between the $17/2^+$ and $13/2^+$ excited states. The corresponding shell gap at $N = 258$ is in agreement with mean-field results (see, e.g., Ref.~\cite{afanasjev2015}). Although this gap is comparable to what we found in $^{209}$Pb at $N=184$, higher-order correlations in the PA-EOM-CCSD expansion could affect this result. While 1p-0h excitations account for more than $90\%$ of the norm of most of the low-lying levels of $^{267}$Pb under consideration, the latter amounts to only $80\%$ for the $13/2^+$ state.

\emph{Conclusions.---} We performed  \textit{ab initio} calculations of $^{208}$Pb and $^{266}$Pb. We validated our theoretical framework by computing the ground-state energy of $^{208}$Pb and by accurately reproducing its first $3^-$ and $2^+$ states. 
We found that the low-lying excited states of $^{266}$Pb are similar to those in $^{208}$Pb, with the $J^\pi=3^-$ being lower in energy than the first $2^+$ state. The energy gap identifies $^{266}$Pb as a doubly magic nucleus. We also found that $^{267}$Pb is unbound with respect to $^{266}$Pb, suggesting that the doubly-magic nucleus is located at the neutron dripline. It is amazing that properties of nuclei around $^{266}$Pb can be predicted with an interaction that was only tuned to light nuclei with mass numbers $A=2,3$ and $4$. 

\emph{Data availability.---} The data that support the findings of this article are openly available~\cite{bonaiti_pb_zenodo2025}.

\acknowledgments This work was supported by the U.S. Department of Energy, Office of Science, Office of Nuclear Physics, under the FRIB Theory Alliance award DE-SC0013617 and award No.~DE-FG02-96ER40963; by the U.S. Department of Energy, Office of Science, Office of Advanced Scientific Computing Research and Office of Nuclear Physics, Scientific Discovery through Advanced Computing (SciDAC) program (SciDAC-5 NUCLEI). This research used resources of the Oak Ridge Leadership Computing Facility located at Oak Ridge National Laboratory, which is supported by the Office of Science of the Department of Energy under contract No. DE-AC05-00OR22725. Computer time was provided by the Innovative and Novel Computational Impact on Theory and Experiment (INCITE) program, by the Institute for Cyber-Enabled Research at Michigan State University, and by the supercomputer Mogon at Johannes Gutenberg Universit\"at Mainz.

\bibliography{master}
\clearpage
\onecolumngrid
\begin{center}
\textbf{Structure of the doubly magic nuclei $^{208}$Pb and $^{266}$Pb from \textit{ab initio} computations: \\Supplemental Material}
\end{center}
\twocolumngrid


This Supplemental Material presents details regarding the convergence of the computed energies with respect to the harmonic oscillator frequency $\hbar\Omega$ and with respect to the three-body force energy cut $E_{\rm 3,max}$, and results for the excited spectra of $^{209,267}$Bi. 

\begin{figure}[htb]
    \centering
    \includegraphics[width=0.49\textwidth]{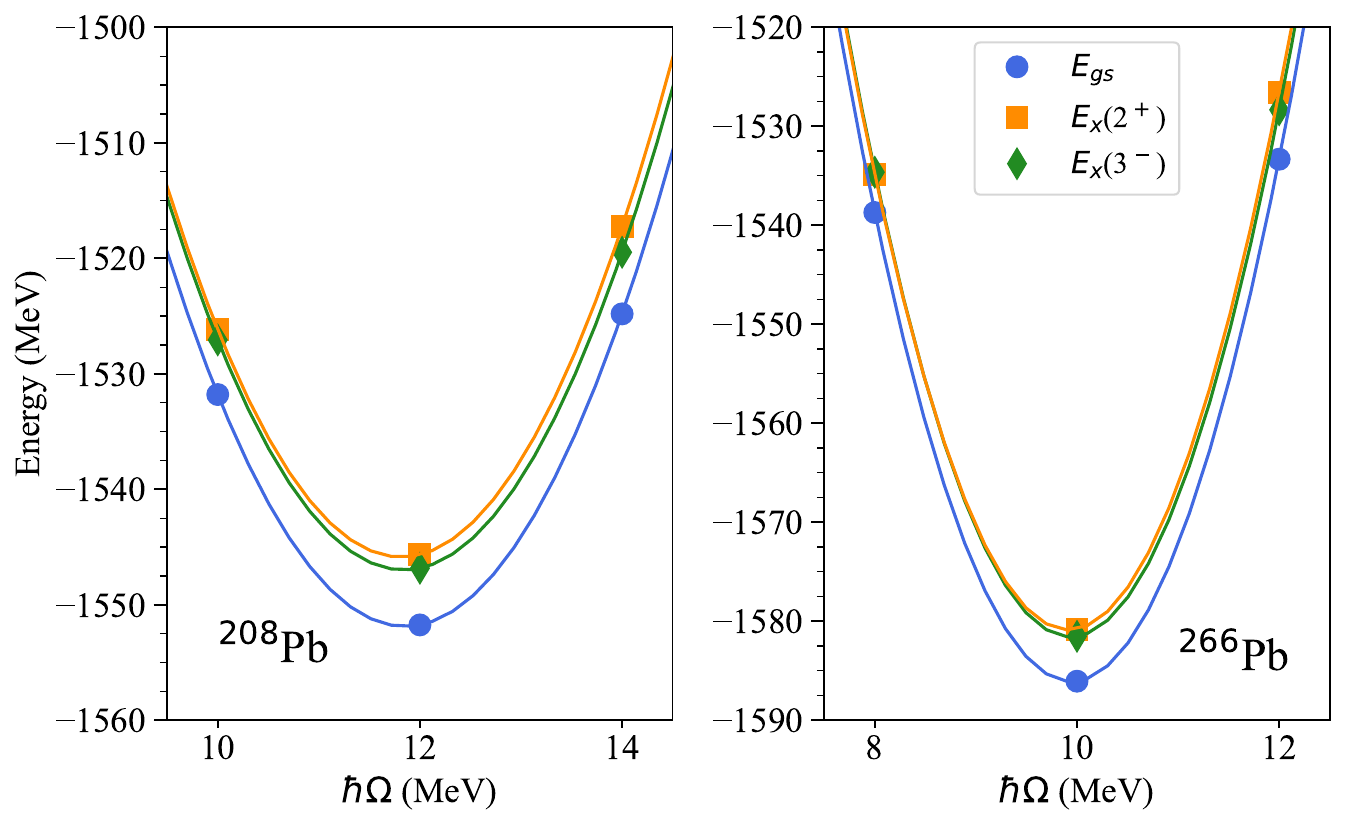}
    \caption{Convergence of the CCSD ground state and first $2^+$ and $3^-$ excited states in $^{208}$Pb (left panel) and $^{266}$Pb (right panel) as a function of the harmonic oscillator frequency $\hbar\Omega$ computed employing the 1.8/2.0 (EM) chiral interaction. Results have been obtained with a model space size $N_{\rm max}$ = 10. Lines correspond to a parabolic fit to the data.}
    \label{fig:hw_dependence}
\end{figure}

\emph{Dependence on $\hbar\Omega$.---} 
Figure~\ref{fig:hw_dependence} shows the ground-state energy and the energy of the first $2^+$ and $3^-$ excited states in $^{208}$Pb and $^{266}$Pb, computed with the CCSD and EOM-CCSD approximation, as a function of the harmonic oscillator frequency $\hbar\Omega$. Due to the considerable computational cost, we performed this study employing a fixed model space size $N_{\rm max} = 10$. The figure also includes a parabolic fit to the result. \\
\\
For both nuclei, the minima of the parabolic curves corresponding to the excited-state results appear to be in close proximity to the ground-state minimum, around $\hbar\Omega = 12$ MeV for $^{208}$Pb and $\hbar\Omega = 10$ MeV for $^{266}$Pb. These are the values employed throughout the main text.

\emph{Convergence with respect to $E_{\rm 3,max}$.---} In Table~\ref{tab:1} (\ref{tab:2}) we report the convergence of the ground- and excited-state energies of $^{208}$Pb ($^{266}$Pb) with respect to the energy cut $E_{\rm 3,max}$ on the three-nucleon force matrix elements. These results have been obtained with $N_{\rm max} = 10$ MeV, $\hbar\Omega = 12$ MeV ($\hbar\Omega = 10$ MeV). 
\begin{table}[htbp]
  \caption{\label{tab:1}{Convergence of the CCSD ground- and excited-state energies of $^{208}$Pb with respect to the three-nucleon force energy cut $E_{\rm 3,max}$ at $N_{\rm max} = 10$, $\hbar\Omega = 12$ MeV. Energies are given in MeV.}}
  \begin{ruledtabular}
    \begin{tabular}{l c c c}
      $E_{\rm 3,max}/\hbar\Omega$ & $E_{g.s.}$  & $E_x(2^+)$ & $E_x(3^-)$ \\ 
      \hline
      24 & -1551.2456 & 6.0637 & 4.8915  \\
      28 & -1551.7436 & 6.0795 & 4.9077 \\
      30 & -1551.7436 & 6.0795 & 4.9077 \\
    \end{tabular}
  \end{ruledtabular}
\end{table}

\begin{table}[htbp]
  \caption{\label{tab:2}{Convergence of the CCSD ground- and excited-state energies of $^{266}$Pb with respect to the three-nucleon force energy cut $E_{\rm 3,max}$ at $N_{\rm max} = 10$, $\hbar\Omega = 10$ MeV. Energies are given in MeV.}}
  \begin{ruledtabular}
    \begin{tabular}{l c c c}
      $E_{\rm 3,max}/\hbar\Omega$ & $E_{g.s.}$  & $E_x(2^+)$ & $E_x(3^-)$ \\ 
      \hline
      24 & -1586.0782 & 5.2567 & 4.5085 \\
      28 & -1586.0527 & 5.2687 & 4.5280 \\
      30 & -1586.0527 & 5.2687 & 4.5280 \\
    \end{tabular}
  \end{ruledtabular}
\end{table}
The numerical values clearly show that the computed energies are fully converged for both nuclei at an energy cut of $E_{\rm 3,max} = 28$, which we employ in the main text. We also point out that the differences between $E_{\rm 3,max} = 24$ and $E_{\rm 3,max} = 28$ are below $0.5\%$ for excitation energies and below $0.05\%$ for ground-state energies.

\begin{figure}[htb]
    \centering
    \includegraphics[width=0.49\textwidth]{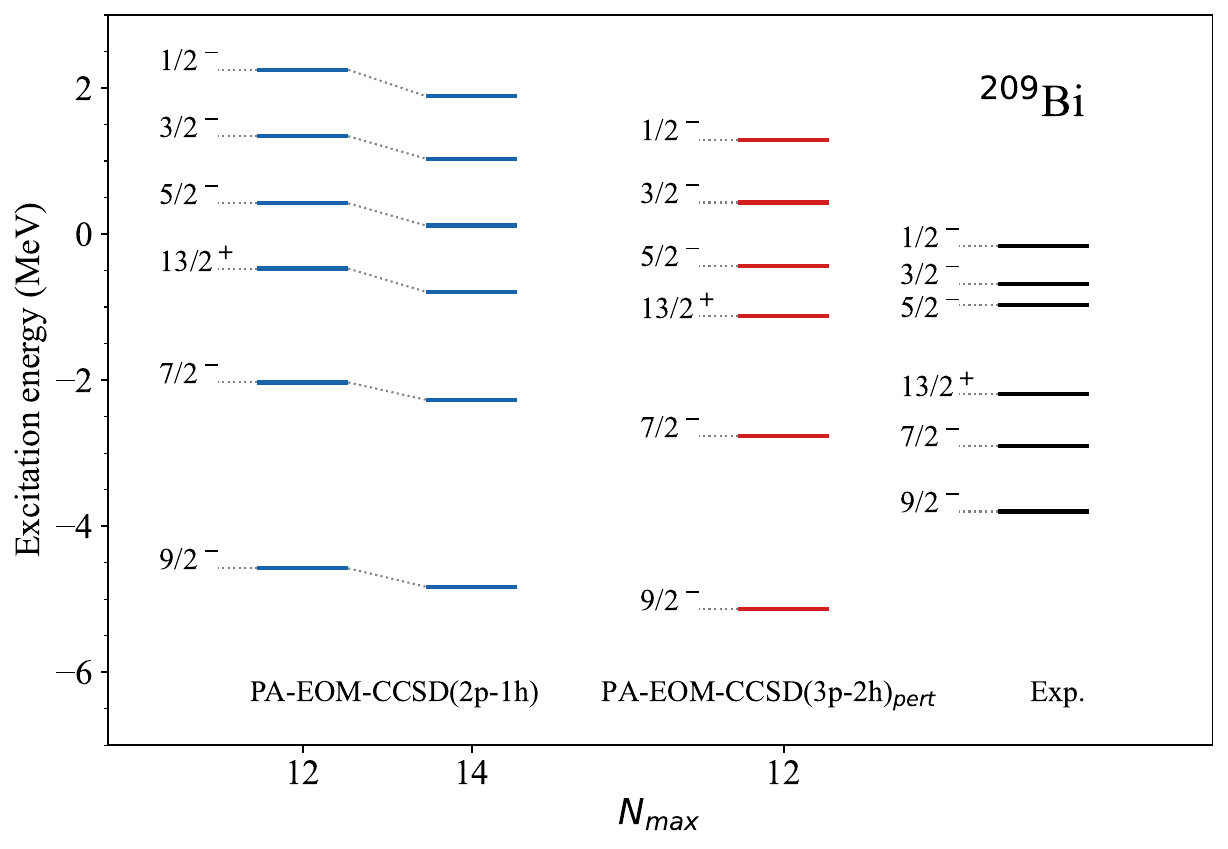}
    \caption{Excited states of $^{209}$Bi computed with the PA-EOM-CCSD(2p-1h) and PA-EOM-CCSD(3p-2h)$_{\rm pert}$ approaches employing the 1.8/2.0 (EM) chiral interaction. Comparison with the experimental data of Ref.~\cite{nudat} is also shown.}
    \label{fig:spectrum_Bi209}
\end{figure}

\emph{Excited states of $^{209,267}$Bi.--} Our PA-EOM-CCSD results for $^{209}$Bi are shown in Fig.~\ref{fig:spectrum_Bi209} in comparison to experimental data~\cite{nudat}. All energies are with respect to the ground state of $^{208}$Pb.

Although the spread of the data is compressed in comparison to the theoretical values, the ordering of the states in $^{209}$Bi is correctly reproduced. Our theoretical results for the excited states of $^{209}$Bi suggest the possible presence of a sub-shell closure based on the $9/2^-$ orbital at $Z=92$, $N=126$. However, large-scale shell model calculations disfavour such a scenario in $^{218}$U~\cite{caurier2003}, in line with the experimental finding of a low-lying $8^+$ isomer at $2.105(20)$ MeV~\cite{leppanen2007}. 
No further gaps are found in the energy spectrum of $^{209}$Bi, suggesting a progressive fading of shell structure for high values of $Z$ (“shell diffusion”), also observed in mean-field calculations~\cite{jerabek2018,giuliani2019}. 

\begin{figure}[htb]
    \centering
    \includegraphics[width=0.48\textwidth]{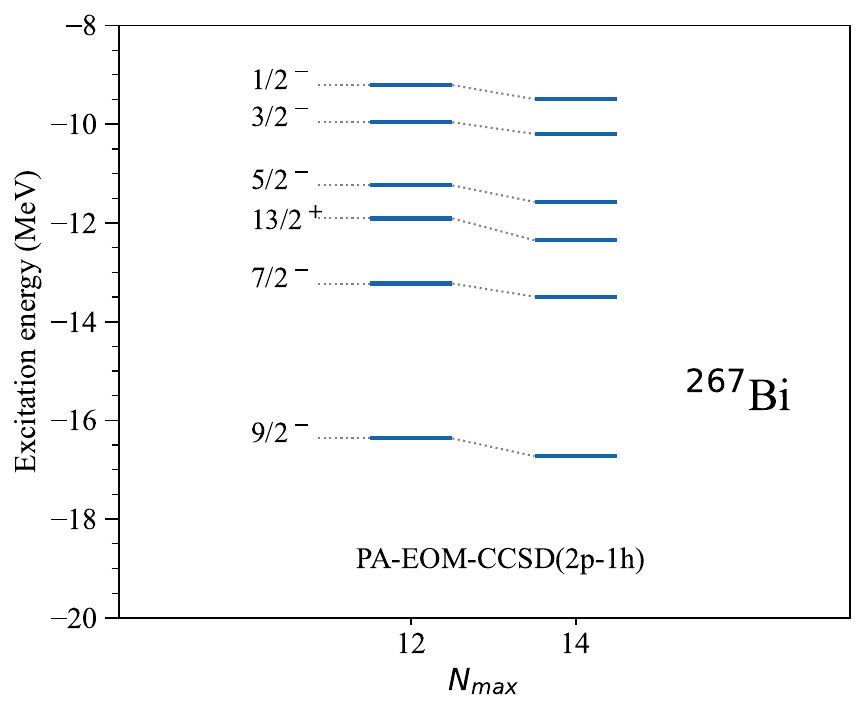}
    \caption{Excited states of $^{267}$Bi computed with the PA-EOM-CCSD(2p-1h) approach employing the 1.8/2.0(EM) chiral interaction.}
    \label{fig:spectrum_Bi267}
\end{figure}

Low-lying states in $^{267}$Bi are shown in Fig.~\ref{fig:spectrum_Bi267}. The proton separation energy of $^{267}$Bi ($N/Z\approx 2.2)$ is $S_p\approx 16$~MeV. These results are typical for neighbors of very neutron-rich  doubly magic nuclei with $N/Z\approx 2$. The data is $S_p\approx 15$~MeV in $^{79}$Cu (neighbor of the $N/Z\approx 1.8$ nucleus $^{78}$Ni) and $S_p\approx 14$~MeV in $^{25}$F (neighbor of the $N/Z = 2$ nucleus $^{24}$O). We note that excited states in $^{209}$Bi and $^{267}$Bi are characterized by the same ordering, consistent with the fact that both nuclei share the same proton shell structure. Again, no clear gap in the energy spectrum can be identified. 
\\


\end{document}